\documentclass[useAMS,usenatbib,letters]{mn2e}
\usepackage{graphicx}

\newcommand{\numax}{$\nu_{\rm max}$}
\newcommand{\logg}{$\log g$}
\newcommand{\teff}{$T_{\rm eff}$}

\title[Assessing the accuracy of the gravity determination in late-type stars with solar-like pulsators]{Assessing the accuracy of the surface gravity determination in late-type stars with solar-like pulsators}

\author[T. Morel and A. Miglio]{Thierry Morel$^{1}$ and Andrea Miglio$^{2}$\\
$^{1}$ Institut d'Astrophysique et de G\'eophysique, Universit\'e de Li\`ege, All\'ee du 6 Ao\^ut, B\^at. B5c, 4000 Li\`ege, Belgium\\
$^{2}$ School of Physics and Astronomy, University of Birmingham, Edgbaston, Birmingham, B15 2TT, UK}
\begin{document}

\date{Accepted 2011 October 10. Received 2011 October 10; in original form 2011 September 7}

\pagerange{\pageref{firstpage}--\pageref{lastpage}} \pubyear{2011}

\maketitle

\label{firstpage}

\begin{abstract}
The frequency of maximum oscillation power measured in dwarfs and giants exhibiting solar-like pulsations provides a precise, and potentially accurate, inference of the stellar surface gravity. An extensive comparison for about 40 well-studied pulsating stars with gravities derived using classical methods (ionisation balance, pressure-sensitive spectral features or location with respect to evolutionary tracks) supports the validity of this technique and reveals an overall remarkable agreement with mean differences not exceeding 0.05 dex (although with a dispersion of up to $\sim$0.2 dex). It is argued that interpolation in theoretical isochrones may be the most precise way of estimating the gravity by traditional means in nearby dwarfs. Attention is drawn to the usefulness of seismic targets as benchmarks in the context of large-scale surveys.
\end{abstract}


\section{Solar-like oscillations as a powerful gravity indicator}
It is notoriously difficult to accurately estimate the stellar surface gravity in late-type stars, with systematic differences of the order of 0.2 dex being commonplace depending on the technique used and its exact implementation. This large uncertainty surrounding \logg \ limits the accuracy with which elemental abundances can be determined. This is especially the case for purely spectroscopic analyses where the determinations of the stellar parameters are intimately coupled. In such a case, the use of a model atmosphere with an inappropriate gravity adversely impacts on the estimation of the other parameters (i.e., effective temperature and microturbulence) and, ultimately, chemical abundances. 

However, the properties of the $p$-mode pulsations exhibited by cool stars on the main sequence and during the red-giant phase can be used to derive values that are precise to a level rivaling that obtained for eclipsing binaries. Although seismic gravities can also be derived using oscillation frequencies and frequency separations, here we only consider the frequency of maximum power, \numax, as a surface gravity indicator (see, e.g., \citealt{kallinger10a} for definition and further details on how this quantity can be derived). As first suggested by \citet{brown} and recently discussed from a theoretical viewpoint by \citet{belkacem}, \numax \ is expected to scale as the acoustic cut-off frequency:

\begin{equation}
{\nu_{\rm max} \over \nu_{{\rm max}, \, \odot}} = 
\left({M\over {\rm M}_\odot}\right) 
\left({R\over {\rm R}_\odot}\right)^{-2} 
\left({T_{\rm eff}\over {\rm T}_{{\rm eff}, \, \odot}}\right)^{-1/2}
\end{equation}

This leads to:

\begin{equation}
\log g = \log g_\odot + \log \left(
{{\nu_{\rm max} \over \nu_{{\rm max}, \, \odot}}}
\right)
+ {1 \over 2}
\log \left({T_{\rm eff}\over {\rm T}_{{\rm eff}, \, \odot}}\right) {\rm .}
\label{eq:loggnumax}
\end{equation}

This relation is largely insensitive to the effective temperature assumed ($\Delta T_{\rm eff}$ = 100 K leads to $\Delta \log g$ $\sim$ 0.004 dex only for Sun-like stars). On the other hand, \numax \ can usually be measured with an error below 5 per cent from high-quality time series (e.g., \citealt{kallinger10a}; \citealt{mosser10}). It follows that \logg \  determined via equation (\ref{eq:loggnumax}) can be precise to better than 0.03 dex. If confirmed in terms of accuracy, this would be far better than what can be achieved by other means in single stars (except in stars with transiting planets, although this method heavily relies on evolutionary models; e.g., \citealt{torres}). Indeed, seismic gravities are beginning to be adopted in spectroscopic analyses as an alternative to values derived from traditional methods in order to narrow down the uncertainties in the other fundamental stellar parameters and chemical abundances (e.g., \citealt{batalha}).

The high accuracy of the gravities obtained from asteroseismology is supported by a comparison with values estimated using completely independent techniques (e.g., as shown in the case of a few binaries by \citealt{bruntt10}, as well as in red-giant cluster members by \citealt{stello11} and \citealt{miglio_etal11}). However, the validity of the scalings relating the stellar parameters (mass, radius) and the seismic observables has yet to be thoroughly investigated for stars occupying different parts of the HR diagram and having various properties in terms of metallicity and activity level, for instance. This work is an effort towards this goal (see also \citealt{miglio11}) and also aims at drawing attention to the usefulness of seismic targets for validation purposes in the context of large-scale stellar surveys.

\section{A sample of well-studied stars with a precise seismic gravity}
About 40 bright, well-studied solar-like and red-giant stars have an accurate estimate of the frequency of maximum power, either from ground-based radial-velocity monitoring or from ultra-precise photometric observations from space (Table \ref{table_numax}). These have been used, along with mean literature \teff \ values (see below) and assuming $\nu_{{\rm max}, \, \odot}$ = 3100 $\umu$Hz, to compute the seismic gravities (the exact choice of $\nu_{{\rm max}, \, \odot}$, which has an uncertainty of $\sim$50 $\umu$Hz, has a negligible impact on our results). The temperatures adopted are marginally higher than those derived from angular diameter and bolometric flux measurements \citep{bruntt10}: $<$$\Delta T_{\rm eff}$$>$=+44$\pm$56 K (1$\sigma$, 10 stars). Adopting these values would lead to negligible differences in the seismic \logg \ (well below 0.01 dex). The uncertainty in \numax, which is the main source of error, is often not quoted in the original literature source or its estimation relies on widely different criteria and assumptions. It is therefore impossible to properly account for the star-to-star differences in the data quality and provide a homogeneous set of uncertainties. Adopting various procedures for the determination of \numax \ and taking into account the different signal realisations arising from the stochastic nature of the oscillations, \citet{hekker} inferred an uncertainty in the range 1--10 per cent for stars observed by the {\it Kepler} mission. Based on the type of data collected for the stars in our sample, we estimate a typical uncertainty of 5 per cent. This translates into an error in the seismic gravities of $\sim$0.03 dex only. These figures are supported by a comparison with the values for the three stars in binaries with dynamical masses and interferometric radii (Procyon A and $\alpha$ Cen A+B; \citealt{bruntt10}): the gravities agree to within 0.02 dex. 

\begin{table*}
\caption{Values of the frequency of maximum power, \numax, for the stars in our sample. The typical uncertainty is 5 per cent (see text). The original references for the seismic data are given. When not explicitly quoted in these papers, the \numax \ values were taken from \citet{bruntt10}, \citet{kallinger10a} or \citet{mosser10}. The value for 18 Sco was computed from the original data.} 
\hspace*{-1cm}
\begin{tabular}{lccl|lccl}
\hline
                               & Name   & \numax \ [$\umu$Hz] &  Reference & & Name & \numax \ [$\umu$Hz] &  Reference\\ 
\hline
\multicolumn{4}{c|}{\bf Dwarfs}                                    &  HD 165341 & 70 Oph A           & 4500     & \citet{carrier_eggenberger}\\
HD    2151 & $\beta$ Hyi        & 1000   &  \citet{kjeldsen}       &  HD 170987 &                    &  930     & \citet{mathur}\\
HD   10700 & $\tau$ Cet         & 4490   &  \citet{teixeira}       &  HD 175726 &                    & 2000     & \citet{mosser09}\\
HD   17051 & $\iota$ Hor        & 2700   &  \citet{vauclair}       &  HD 181420 &                    & 1500     & \citet{barban09}\\
HD   20010 & $\alpha$ For       & 1100   &  \citet{kjeldsen}       &  HD 181906 &                    & 1912     & \citet{garcia}\\
HD   23249 & $\delta$ Eri       &  700   &  \citet{bouchy_carrier} &  HD 190248 & $\delta$ Pav       & 2300     & \citet{kjeldsen}\\
HD   49385 &                    & 1013   &  \citet{deheuvels}      &  HD 203608 & $\gamma$ Pav       & 2600     & \citet{mosser08}\\
HD   49933 &                    & 1657   &  \citet{kallinger10b}   &  HD 210302 & $\tau$ Psa         & 1950     & \citet{bruntt10}\\
HD   52265 &                    & 2090   &  \citet{ballot}         &            &                    &          &   \\
HD   61421 & Procyon A          & 1000   &  \citet{arentoft}       & \multicolumn{4}{c}{\bf Subgiants and giants} \\
HD   63077 & 171 Pup            & 2050   &  \citet{bruntt10}       &  HD  71878 & $\beta$ Vol        &   51     & \citet{stello09}\\
HD  102870 & $\beta$ Vir        & 1400   &  \citet{carrier05a}     &  HD 100407 & $\xi$ Hya          &   92.3   & \citet{frandsen}\\
HD  121370 & $\eta$ Boo         &  750   &  \citet{carrier05b}     &  HD 124897 & $\alpha$ Boo       &    3.47  & \citet{tarrant})\\
HD  128620 & $\alpha$ Cen A     & 2400   &  \citet{kjeldsen}       &  HD 146791 & $\epsilon$ Oph     &   53.5   & \citet{kallinger08}\\
HD  128621 & $\alpha$ Cen B     & 4100   &  \citet{kjeldsen}       &  HD 153210 & $\kappa$ Oph       &   35     & \citet{stello09}\\
HD  139211 & HR 5803            & 2800   &  \citet{carrier08}      &  HD 161096 & $\beta$ Oph        &   46     & \citet{kallinger10a}\\
HD  142860 & $\gamma$ Ser       & 1600   &  \citet{kjeldsen}       &  HD 163588 & $\xi$ Dra          &   36     & \citet{stello09}\\
HD  146233 & 18 Sco             & 3170   &  \citet{bazot}          &  HD 168723 & $\eta$ Ser         &  125     & \citet{barban04}\\
HD  150680 & $\zeta$ Her A      &  700   &  \citet{martic}         &  HD 188512 & $\beta$ Aql        &  410     & \citet{kjeldsen}\\
HD  160691 & $\mu$ Ara          & 2000   &  \citet{bouchy}         &  HD 211998 & $\nu$ Ind          &  313     & \citet{bedding}\\
HD  161797 & $\mu$ Her          & 1200   &  \citet{bonanno}        &            & M67 S1305          &  208.9   & \citet{kallinger10a}\\
\hline
\end{tabular}
\label{table_numax}
\end{table*}

\section{The classical gravity diagnostics used in cool stars put to the test}
As the stars in Table \ref{table_numax} are amongst the brightest in the sky and are even sometimes regarded as standards (e.g., $\alpha$ Boo or Procyon A), a large number of independent determinations from classical techniques can be found in the literature. This offers an opportunity to empirically assess the reliability of the most popular gravity diagnostics used in cool stars: ionisation balance of a given chemical species (usually iron), fitting the wings of pressure-sensitive, neutral metal lines or interpolation in theoretical isochrones. Discrepancies between the various indicators are known to exist in the case of unresolved, single-lined binaries \citep{fuhrmann}, but this should not be a concern in our sample. These three approaches suffer to different extents from drawbacks. First, the values obtained from ionisation equilibrium are strongly dependent on the atmospheric structure adopted (granulation is ignored in 1-D models) and can be biased by non-LTE effects, which become generally more important in stars with extended atmospheres and/or metal poor (see, e.g., the calculations of \citealt{mashonkina} applied to some stars in our sample). On the other hand, values obtained from fitting the wings of pressure-sensitive lines are generally affected by quite large uncertainties related, for instance, to blends and difficulties in continuum placement (e.g., \citealt{bruntt10}). Finally, although for very nearby stars parallaxes and reddening are not a major concern, interpolation in theoretical isochrones is strongly model dependent and may suffer from degeneracy problems (as a result, the applicability of this method for stars on the red-giant branch is limited).

The \teff, [Fe/H] and \logg \ literature values for the stars in Table \ref{table_numax} were primarily extracted from the PASTEL catalogue \citep{soubiran}, but were supplemented by data from several missing sources after a careful inspection of the
vast literature for these objects. Only studies published after 1990 were considered, as older ones may be based on poor-quality data or inadequate model atmospheres. Each original reference was inspected to evaluate the method used for the \logg \ determination. In some instances, a single value was quoted in PASTEL whereas estimates based on different techniques were reported in the original paper (e.g., \citealt{santos05} where the gravities estimated from isochrone fitting are missing). These values were added. Finally, duplicate entries from the same authors were omitted; only the value in the most recent paper was used. This roughly totals to 360 individual measurements from 80 independent literature sources. The results are presented in Table \ref{table_parameters}. 

\begin{table*}
\centering
\caption{Mean effective temperature, iron content and mean surface gravities from the four different methods for the stars in Table \ref{table_numax}. The error bars are the quadratic sum of the standard deviation of the individual measurements and the typical uncertainty in the parameter determination (80 K for \teff, 0.1 dex for [Fe/H], 0.1 dex for the ionisation and isochrone gravities, and 0.15 dex for the strong-line gravities). The numbers in brackets are the number of measurements. The typical uncertainty in the seismic \logg \ is 0.03 dex (see text). } 
\begin{tabular}{lccccccc}
\hline
         &                &                   &                     & \multicolumn{4}{c}{\logg}                                        \\ 
         & Name           & \teff \ [K] &  [Fe/H]             & seismology & ionisation         & wings              & isochrone        \\ 
\hline
\multicolumn{8}{c}{\bf Dwarfs}\\
HD   2151 & $\beta$ Hyi    & 5829$\pm$107 (10) &  --0.09$\pm$0.12 (11) & 3.95 & 4.02$\pm$0.18 (6)  &  3.76$\pm$0.15 (1) &  3.98$\pm$0.10 (8)\\
HD  10700 & $\tau$ Cet     & 5334$\pm$103 (14) &  --0.53$\pm$0.11 (14) & 4.58 & 4.48$\pm$0.21 (12) &  4.45$\pm$0.15 (1) &  4.51$\pm$0.14 (5)\\
HD  17051 & $\iota$ Hor    & 6136$\pm$120  (9) &    0.15$\pm$0.12  (8) & 4.39 & 4.48$\pm$0.15 (6)  &  4.40$\pm$0.15 (1) &  4.40$\pm$0.13 (5)\\
HD  20010 & $\alpha$ For   & 6154$\pm$141  (7) &  --0.26$\pm$0.12  (7) & 4.00 & 4.07$\pm$0.25 (4)  &  3.79$\pm$0.15 (1) &  3.97$\pm$0.11 (4)\\
HD  23249 & $\delta$ Eri   & 5060$\pm$111 (11) &    0.12$\pm$0.13 (11) & 3.76 & 3.86$\pm$0.18 (5)  &  3.95$\pm$0.27 (2) &  3.82$\pm$0.21 (5)\\
HD  49385 &                & 6131$\pm$94   (2) &    0.09$\pm$0.10  (1) & 3.97 & 4.00$\pm$0.10 (1)  &  4.03$\pm$0.15 (1) &  4.08$\pm$0.10 (1)\\
HD  49933 &                & 6580$\pm$120  (7) &  --0.44$\pm$0.10  (5) & 4.20 & 4.24$\pm$0.19 (4)  &  4.00$\pm$0.15 (1) &  4.23$\pm$0.14 (3)\\
HD  52265 &                & 6097$\pm$92  (12) &    0.19$\pm$0.11 (11) & 4.28 & 4.31$\pm$0.16 (9)  &                    &  4.29$\pm$0.11 (6)\\
HD  61421 & Procyon A      & 6590$\pm$131 (13) &  --0.03$\pm$0.11 (15) & 3.98 & 4.06$\pm$0.32 (9)  &  3.92$\pm$0.20 (2) &  4.01$\pm$0.11 (8)\\
HD  63077 & 171 Pup        & 5783$\pm$135  (8) &  --0.86$\pm$0.14  (6) & 4.26 & 4.16$\pm$0.19 (3)  &  4.00$\pm$0.15 (1) &  4.22$\pm$0.15 (5)\\
HD 102870 & $\beta$ Vir    & 6131$\pm$107 (11) &    0.13$\pm$0.11 (11) & 4.11 & 4.11$\pm$0.16 (7)  &  3.97$\pm$0.15 (1) &  4.13$\pm$0.11 (7)\\
HD 121370 & $\eta$ Boo     & 6059$\pm$143  (9) &    0.23$\pm$0.11  (9) & 3.83 & 3.83$\pm$0.29 (7)  &  3.90$\pm$0.15 (1) &  3.80$\pm$0.11 (4)\\
HD 128620 & $\alpha$ Cen A & 5745$\pm$138 (14) &    0.21$\pm$0.13 (14) & 4.33 & 4.21$\pm$0.21 (9)  &  4.32$\pm$0.15 (1) &  4.31$\pm$0.11 (6)\\
HD 128621 & $\alpha$ Cen B & 5191$\pm$126  (9) &    0.24$\pm$0.11  (9) & 4.54 & 4.46$\pm$0.12 (5)  &  4.52$\pm$0.15 (1) &  4.54$\pm$0.11 (5)\\
HD 139211 &                & 6296$\pm$161  (3) &  --0.15$\pm$0.18  (2) & 4.41 & 4.05$\pm$0.10 (1)  &  4.10$\pm$0.15 (1) &  4.20$\pm$0.15 (2)\\
HD 142860 & $\gamma$ Ser   & 6253$\pm$108 (10) &  --0.19$\pm$0.12 (10) & 4.17 & 4.05$\pm$0.17 (6)  &  4.02$\pm$0.15 (1) &  4.20$\pm$0.12 (6)\\
HD 146233 & 18 Sco         & 5783$\pm$92  (13) &    0.03$\pm$0.11 (13) & 4.45 & 4.40$\pm$0.13 (10) &                    &  4.43$\pm$0.11 (7)\\
HD 150680 & $\zeta$ Her A  & 5762$\pm$110  (6) &    0.01$\pm$0.12  (6) & 3.79 & 3.85$\pm$0.18 (3)  &                    &  3.71$\pm$0.11 (5)\\
HD 160691 & $\mu$ Ara      & 5732$\pm$104 (12) &    0.26$\pm$0.11 (11) & 4.25 & 4.20$\pm$0.20 (6)  &  4.07$\pm$0.15 (1) &  4.23$\pm$0.11 (9)\\
HD 161797 & $\mu$ Her      & 5532$\pm$105  (7) &    0.23$\pm$0.13  (7) & 4.02 & 3.98$\pm$0.10 (3)  &                    &  3.94$\pm$0.17 (5)\\
HD 165341 & 70 Oph A       & 5221$\pm$135  (7) &    0.00$\pm$0.15  (7) & 4.58 & 4.38$\pm$0.19 (5)  &  4.56$\pm$0.15 (1) &  4.52$\pm$0.11 (3)\\
HD 170987 &                & 6540$\pm$80   (1) &  --0.15$\pm$0.10  (1) & 3.94 &                    &  4.35$\pm$0.15 (1) &                   \\
HD 175726 &                & 6031$\pm$88   (3) &  --0.07$\pm$0.10  (2) & 4.26 & 4.53$\pm$0.10 (1)  &                    &  4.38$\pm$0.10 (2)\\
HD 181420 &                & 6671$\pm$151  (2) &    0.00$\pm$0.10  (1) & 4.15 & 4.26$\pm$0.10 (1)  &                    &  4.23$\pm$0.10 (1)\\
HD 181906 &                & 6607$\pm$80   (1) &                       & 4.26 &                    &                    &  4.24$\pm$0.10 (1)\\
HD 190248 & $\delta$ Pav   & 5558$\pm$129  (9) &    0.30$\pm$0.16  (9) & 4.30 & 4.23$\pm$0.15 (5)  &  4.32$\pm$0.15 (1) &  4.32$\pm$0.12 (6)\\
HD 203608 & $\gamma$ Pav   & 6065$\pm$109 (11) &  --0.73$\pm$0.13 (10) & 4.37 & 4.22$\pm$0.35 (4)  &  4.15$\pm$0.15 (1) &  4.33$\pm$0.12 (7)\\
HD 210302 & $\tau$ Psa     & 6295$\pm$96   (3) &    0.05$\pm$0.11  (2) & 4.26 & 4.09$\pm$0.10 (1)  &  4.11$\pm$0.15 (1) &  4.25$\pm$0.12 (2)\\
\multicolumn{8}{c}{\bf Subgiants and giants}\\
HD  71878 & $\beta$ Vol    & 4736$\pm$246  (2) &  --0.01$\pm$0.10  (1) & 2.61 & 3.00$\pm$0.10 (1)  &                    &  2.42$\pm$0.10 (1)\\
HD 100407 & $\xi$ Hya      & 5002$\pm$106  (5) &    0.11$\pm$0.15  (4) & 2.88 & 2.86$\pm$0.17 (2)  &  2.88$\pm$0.15 (1) &  2.69$\pm$0.23 (3)\\
HD 124897 & $\alpha$ Boo   & 4292$\pm$97  (17) &  --0.58$\pm$0.11 (17) & 1.42 & 1.61$\pm$0.25 (11) &                    &  1.84$\pm$0.29 (7)\\
HD 146791 & $\epsilon$ Oph & 4921$\pm$98   (8) &  --0.09$\pm$0.13  (7) & 2.64 & 2.82$\pm$0.20 (4)  &                    &  2.73$\pm$0.23 (5)\\
HD 153210 & $\kappa$ Oph   & 4559$\pm$116  (4) &    0.06$\pm$0.13  (3) & 2.44 & 2.50$\pm$0.30 (2)  &                    &  2.47$\pm$0.24 (2)\\
HD 161096 & $\beta$ Oph    & 4580$\pm$112  (6) &    0.14$\pm$0.13  (5) & 2.56 & 2.67$\pm$0.26 (3)  &                    &  2.38$\pm$0.24 (3)\\
HD 163588 & $\xi$ Dra      & 4464$\pm$123  (3) &  --0.05$\pm$0.11  (2) & 2.45 & 2.40$\pm$0.10 (1)  &                    &  2.46$\pm$0.24 (2)\\
HD 168723 & $\eta$ Ser     & 4927$\pm$89  (10) &  --0.18$\pm$0.14  (9) & 3.01 & 3.06$\pm$0.15 (7)  &  2.95$\pm$0.15 (1) &  3.09$\pm$0.13 (6)\\
HD 188512 & $\beta$ Aql    & 5100$\pm$93   (8) &  --0.20$\pm$0.12  (8) & 3.53 & 3.58$\pm$0.14 (3)  &  3.69$\pm$0.15 (1) &  3.55$\pm$0.11 (5)\\
HD 211998 & $\nu$ Ind      & 5244$\pm$101  (7) &  --1.54$\pm$0.14  (6) & 3.42 & 3.31$\pm$0.18 (3)  &  3.70$\pm$0.15 (1) &  3.40$\pm$0.11 (4)\\
          &  M67 S1305     & 4940$\pm$80   (1) &  --0.08$\pm$0.10  (1) & 3.23 & 3.20$\pm$0.10 (1)  &                    &                   \\
\hline
\end{tabular}
\label{table_parameters}
\end{table*}

\begin{figure*}
\includegraphics[width=175mm]{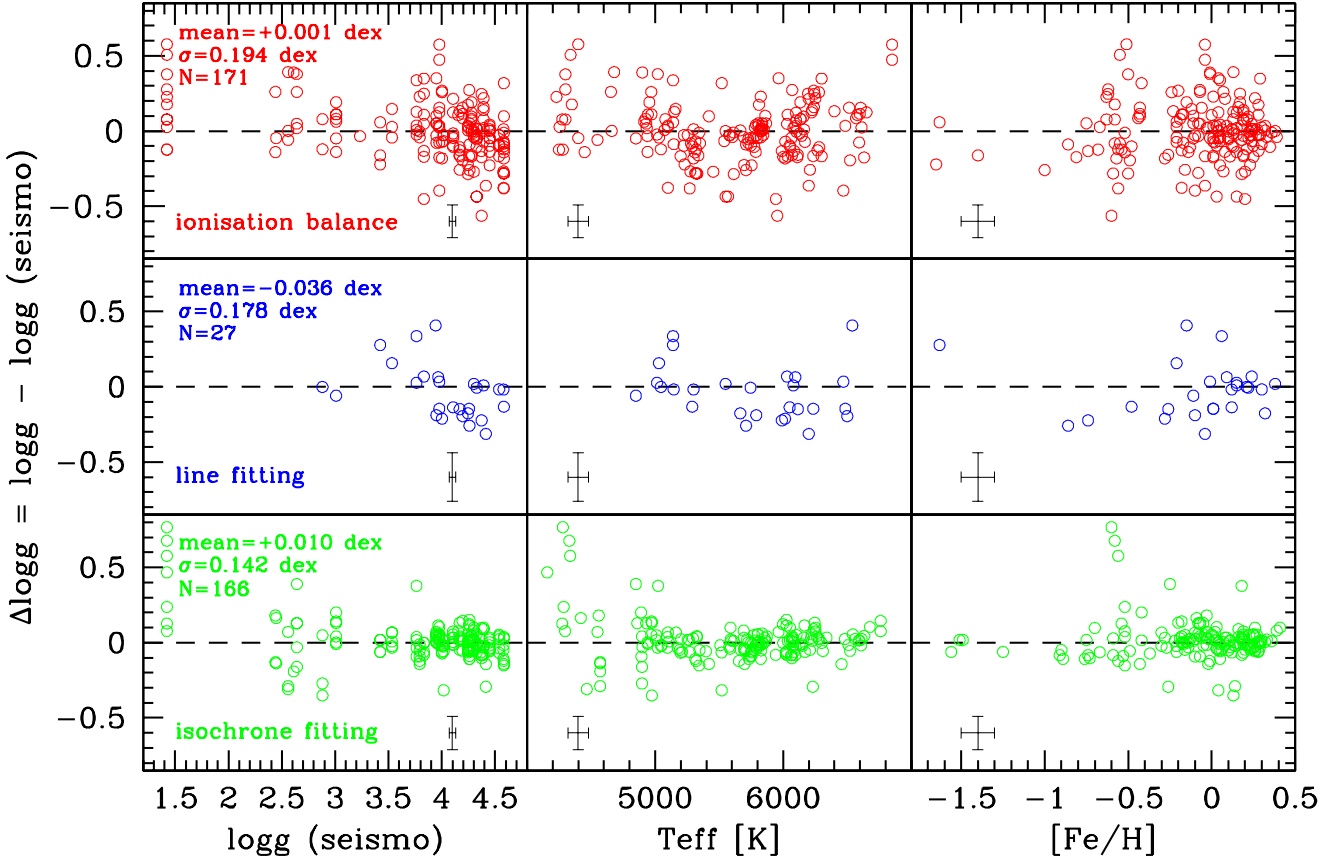}
\caption{Difference between the seismic \logg \ values and those obtained through ionisation balance of iron ({\it top panels}), fitting of the wings of pressure-sensitive lines ({\it middle panels}) and isochrone fitting ({\it bottom panels}), as a function of the seismic gravities, effective temperature and metallicity. Representative error bars are indicated.}
\label{fig_measurements}
\end{figure*}

\begin{figure*}
\includegraphics[width=175mm]{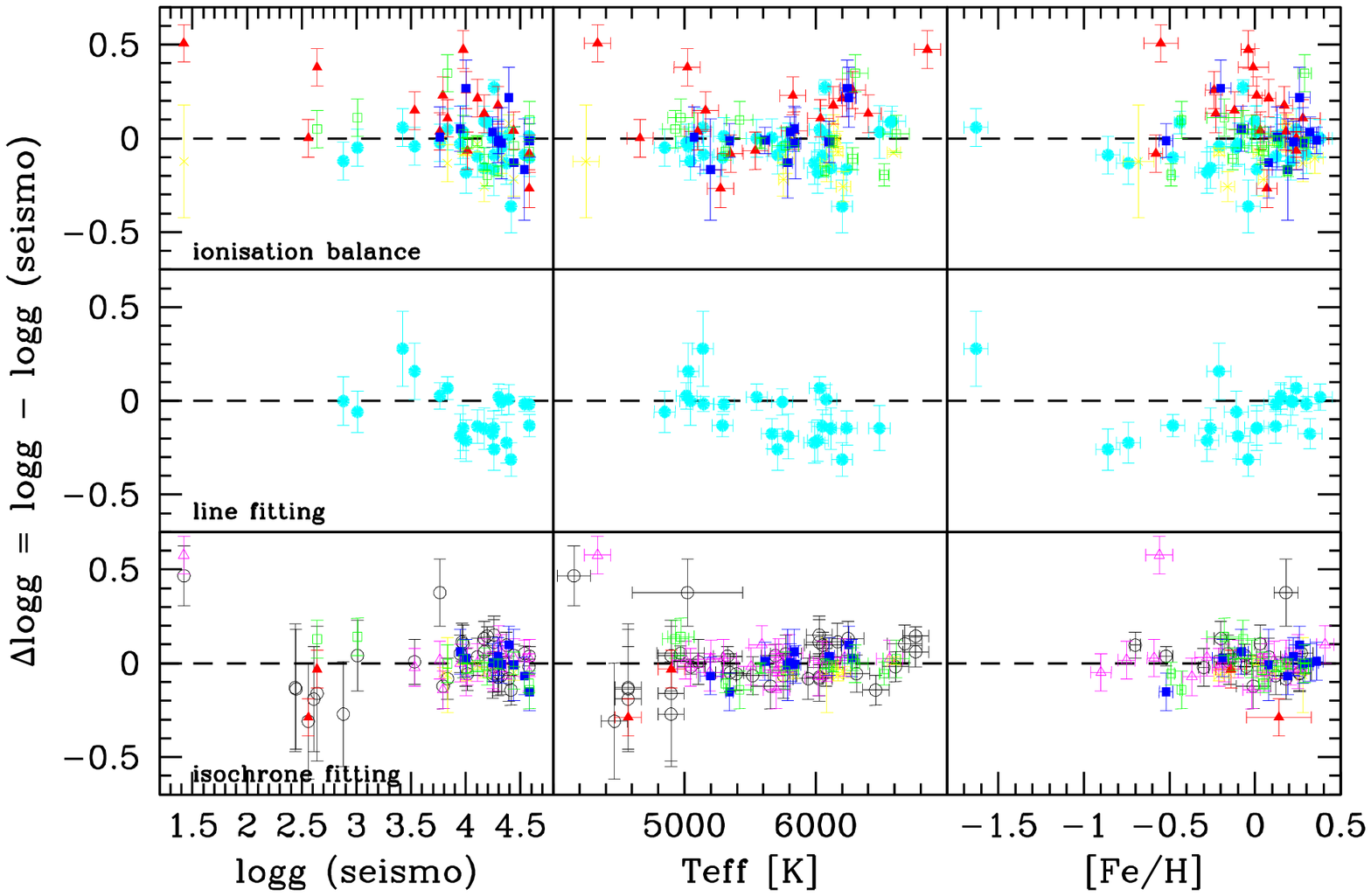}
\caption{As Fig.\ref{fig_measurements}, but for the key studies only. 
{\it Black, open circles}: \citet{allende_lambert} and \citet{allende04};
{\it cyan, filled circles}: \citet{bruntt09} and \citet{bruntt10};
{\it yellow crosses}: \citet{gonzalez_wallerstein}, \citet{gonzalez01} and \citet{gonzalez10};
{\it red, filled triangles}: \citet{luck_heiter06, luck_heiter07};
{\it magenta, open triangles}: \citet{ramirez};
{\it blue, filled squares}: \citet{santos01, santos04} and \citet{santos05};
{\it green, open squares}: \citet{takeda05, takeda07} and \citet{takeda08}.}
\label{fig_studies}
\end{figure*}

The comparison between the seismic \logg \ values and those obtained through traditional techniques is shown in Fig.\ref{fig_measurements}. Overall, there is a remarkably good agreement with systematic differences not exceeding 0.04 dex on average. The significant 1-$\sigma$ dispersion of up to 0.19 dex with respect to the reference seismic value may have been expected considering the heterogeneous nature of the data and the diversity of analyses performed. It remains to be seen, however, if part of the observed scatter is not due to an intrinsic dispersion of equation (2), which is based on a simple scaling of \numax \ with the acoustic cut-off frequency in the atmosphere. By averaging results from a large number of independent studies (as is the case here for the ionisation and isochrone gravities, but {\it not} for the strong-line ones), one can hope that the systematic errors partly cancel out and that the mean offset with respect to the seismic values provides a better appraisal of the true accuracy of the method. It should be kept in mind that the systematic differences which may exist between the various analyses might not be completely related to the technique used, but instead to other assumptions in the modelling (especially the \teff \ scale). The results of the studies with the highest number of measurements are shown in Fig.\ref{fig_studies}. As can be seen, several of them can be in error by more than a factor 2.

\begin{figure*}
\includegraphics[width=175mm]{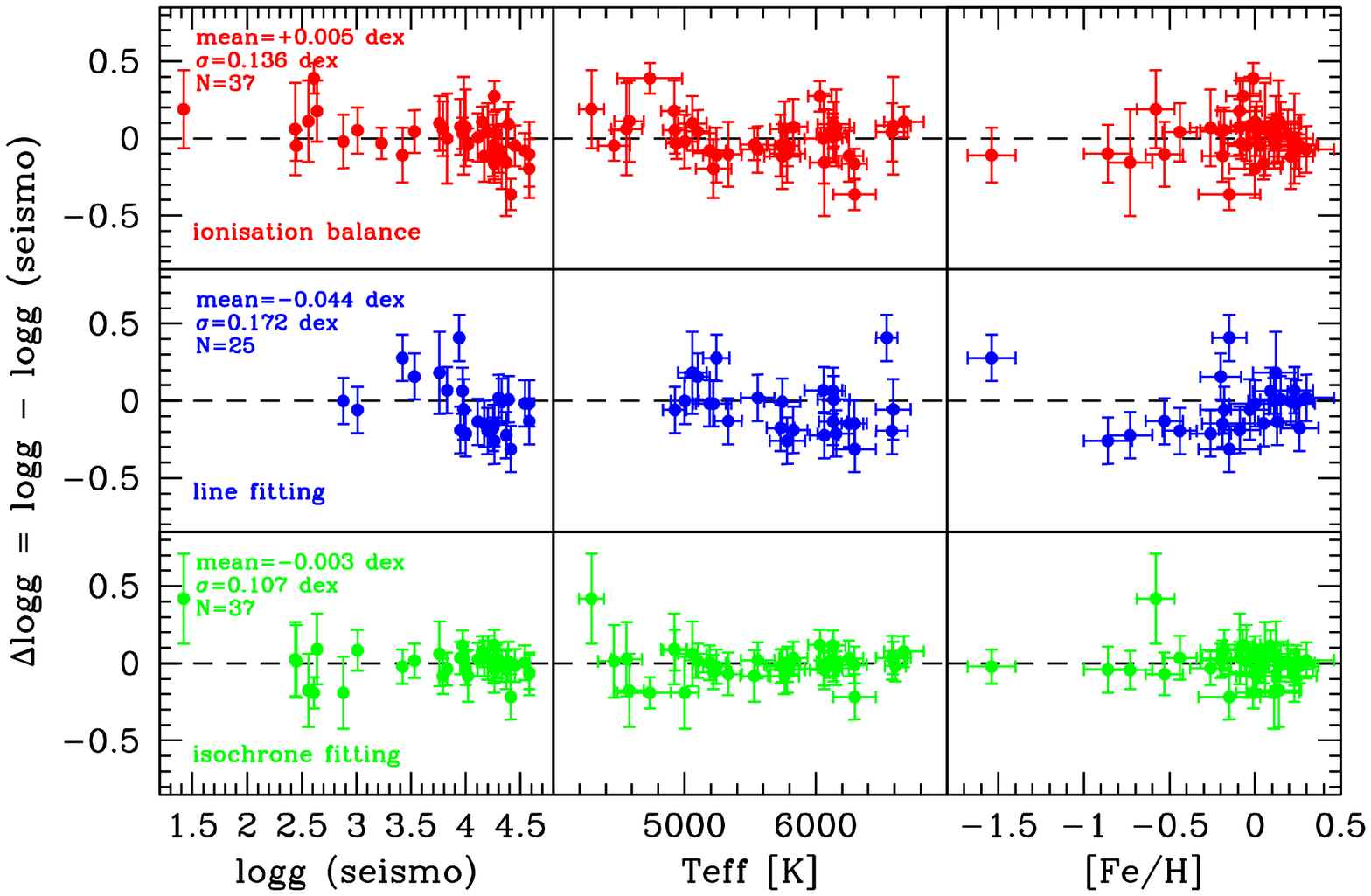}
\caption{As Fig.\ref{fig_measurements}, but with the data averaged on a star-to-star basis.}
\label{fig_stars}
\end{figure*}

It can readily be seen in Fig.\ref{fig_measurements} that the scatter is lower for the gravities estimated from isochrone fitting. The same conclusion holds when considering  for each star the average of the measurements obtained using a given method (Fig.\ref{fig_stars}),\footnote{An unweighted mean has been used because of the difficulty in assessing the quality of the studies performed and the inhomogeneous nature of the quoted uncertainties (which are often only internal and, as a result, appear unrealistically small).} especially when one excludes the evolved objects (\logg \ $<$ 3.2) for which the determination through the position of the star with respect to evolutionary tracks is ill defined. In that case, the difference scatter is a mere $\sim$15 per cent: $<$$\Delta \log g$$>$=--0.006$\pm$0.065 dex (1$\sigma$, 29 stars). The evolutionary tracks used mostly fall in two categories differing in their physical ingredients (e.g., treatment of convection): either those from the Geneva (e.g., \citealt{schaller}) or from the Padova (e.g., \citealt{bertelli}) group. None the less, the gravity determination seems fairly robust against the choice of the set of models adopted (see \citealt{allende99}). Although this method is generally the most precise, it must be stressed that the mean difference with respect to the seismic gravities is less than 0.05 dex irrespective of the technique used.

No clear trends are discernible as a function of \logg, \teff \ or [Fe/H]. An underestimation of \logg \ through ionisation balance may be expected for very metal-poor stars ([Fe/H] $\la$ --1) because of non-LTE effects \citep{allende99}. We only have one such star in our sample ($\nu$ Ind), but the ionisation gravity does not appear discrepant. The \logg \ values are systematically underestimated in the dwarfs by up to 0.3 dex when fitting the wings of strong metal lines. However, the bulk of the data comes from a single source \citep{bruntt10} and large line-to-line differences are observed (a weighted mean has been used here).

\section{Conclusions and perspectives}
The good agreement between the gravities inferred from asteroseismology and from classical methods supports the applicability of the scaling law linking \logg \ and \numax \ for stars spanning a relatively wide range in temperature and evolutionary status, although its validity in the low-metallicity and low-gravity regimes cannot be meaningfully investigated here owing to the limited number of objects. For stars within the parameter space investigated here, our study hence supports the use of the seismic gravities as input in spectroscopic analyses (e.g., \citealt{batalha}). Gravities relying on asteroseismic information would be especially valuable for red giants in view of the fundamental difficulties plaguing the classical techniques in that case (significant departures from LTE, lack of sensitivity of strong metal lines and degeneracy of isochrone fitting). The spectroscopic gravities of the faint red-giant {\it CoRoT} targets are indeed found to be affected by large errors \citep{morel}. Seismic information is now available for hundreds of stars in the {\it CoRoT} and {\it Kepler} fields (see, e.g., \citealt{verner11b}). It is therefore reasonable to assume that similar consistency checks as those presented in this paper will be extended to much larger samples in the future.

A comparison with data for eclipsing binaries has already illustrated the power of isochrone fitting as gravity indicator \citep{allende_lambert}, and our study indeed identifies it as being the most precise classical method for nearby dwarfs. The release of the {\it GAIA} parallaxes will offer the opportunity to apply this technique to much fainter magnitudes, although reddening will remain an issue.

Several large-scale spectroscopic surveys are presently conducted or are being planned (e.g., RAVE, APOGEE or the follow up of the {\it GAIA} mission). The pipelines developed for that purpose should be able to recover the parameters determined for a set of training stars through completely different and, as much as possible, model-independent methods before embarking on the automatic analysis of large samples of potentially faint objects. The seismic gravities can hence constitute a valuable piece of information in this context. Of particular interest in this respect are the stars with an accurate \teff \ and \logg \ estimate from interferometric and seismic observations, respectively (see \citealt{bruntt10}). 

\section*{Acknowledgments}
T. M. acknowledges financial support from Belspo for contract PRODEX GAIA-DPAC and is indebted to S. Vauclair for drawing his attention to the accuracy of the seismic gravities. We would like to thank the anonymous referee for valuable comments, A.-M. Broomhall for recomputing the \numax \ value of 18 Sco and M. Valentini for useful discussions.

\bsp

\label{lastpage}

\end{document}